\documentclass[10pt,letterpaper]{article}
\usepackage{geometry}
\usepackage{amsmath,amssymb}
\usepackage{cite}
\usepackage{color}

\date{}
\usepackage{graphicx}
\usepackage{epstopdf}

\begin{document}

{\Large
\textbf\newline{Basis of Self-organized Proportion Regulation Resulting from Local Contacts}
}
\newline
\\
Mayuko Iwamoto\textsuperscript{1, 2, *},
Daishin Ueyama\textsuperscript{1, 2, \dag}
\\
\newline
1 \  Department of Mathematical Sciences Based on Modeling and Analysis, School of Interdisciplinary Mathematical Sciences, Meiji University, 4-21-1 Nakano, Nakano-ku, Tokyo, Japan
\\
2 \  Meiji Institute for Advanced Study of Mathematical Sciences (MIMS), 4-21-1 Nakano, Nakano-ku, Tokyo, Japan
\\
\bigskip

* miwamoto@meiji.ac.jp

\dag daishin@meiji.ac.jp

\section*{Abstract}

\ \ \ \ One of the fundamental problems in biology concerns the method by which a cluster of organisms can regulate the proportion of individuals that perform various roles or modes as if each individual knows a whole situation without a leader.
A specific ratio exists in various species at multiple levels from the process of cell differentiation in multicellular organisms to the situation of social dilemma in a group of human beings.
This study found a common basis of regulating a collective behavior which is realized by a series of local contacts between individuals. The most essential behavior of individuals in this theory is to change its internal mode through sharing information in contact with others. Our numerical simulations with cellular automata model realize to regulate the ratio of population of individuals who has either two kinds of modes. From the theoretical analysis and numerical calculations, we found that asymmetric properties in local contacts, are essential for adaptive regulation in response to the global information such a group size and whole density. Furthermore, a discrete system is crucial in no-leader groups to realize the flexible regulation, and the critical condition which eliminates overlap with one another (excluded volume effect) also affects the resulting proportion in high density. The foremost advantage of this strategy is that no global information is required for each individual, and only a couple of mode switching can achieve the whole proportion regulation. The simple mechanism say that proportion regulation in well-organized groups in nature can be realized through and limited to local contacts, and has a potential to solve various phenomena that microscopic individuals behaviors connect to macroscopic orderly behaviors. 

\section*{Introduction}
\ \ \ \ 
Learn wisdom by the follies of others.
The same may be equally true of other social animals.
One of the fundamental problems in nature is regulating the proportions of individuals who perform multiple roles or modes within a cluster of organisms. In general, social organizations contain hierarchies and are generally managed through a direct chain of command with a clearly identified superior. Collective organisms found in nature, however, can self-regulate proportions for multiple roles without requiring a central control. For example, the proportion of two different cell types, prestalk and prespore cells, in the slug of the cellular slime mold~\cite{Raper1940,Bonner1957,Bonner1949,Stenhouse-Williams1977,Nanjundiah1995,Rafols2000,Oohata1995}, the proportion of ``active" and ``inactive'' workers~\cite{Prigogine1984,Hasegawa2013,Hayashi2015} and task allocation~\cite{Wilson1971,Gordon1987,Gordon1989,Gordon1996} in ants colony, the proportion of ``aggressive" and ``docile" female spiders~\cite{Pruitt-Goodnight2014}, and the producer-scrounger foraging strategy found in bird crowds~\cite{Giraldeau1994}. 

As reported in the experimental researches on the proportion of some species~\cite{Bonner1949,Stenhouse-Williams1977,Nanjundiah1995,Rafols2000,Oohata1995,Gordon1987,Gordon1989,Gordon1996,Pruitt-Goodnight2014}, these regulating systems may appear a feature to response to the global and temporal properties, e.g., group size and density, which is changed by external disturbances. It is sometimes called ``collective intelligence" due to they can adapt the proportion as if they know the total number of individuals, density for instance. 
Self-organizing proportion regulations in collective behaviors which are universal at multiple levels, such as cellular and individual levels, indicate that the coexistence of multiple modes in suitable proportion is deeply involved in evolution and natural selection.

In addition, the temporal dynamics of internal modes appear in diverse species~\cite{Shimada1995,Viswanathan1996,Nakamura2007,Sims2008}, and it has been reported that each cell or individual in some species can change its mode after the contact with others physically close through specific substances~\cite{Kay1983,Inoue1989,Gordon2011,Hojo2015,Hayashi2015}. 
From the observations that the transdifferentiation of cellular slime molds occurs between two cell types~\cite{Brown-Firtel1999} and the contact that changes tasks in ants is performed between two ants~\cite{Gordon2011}, the models for the mechanisms on proportion regulation, which maintain the idea of local interaction between individuals, have been suggested~\cite{Nanjundiah1995,Gordon1992,Kaneko2013}. However, these models implicitly include a global property, hence it remains a mystery how each individual can know it.

While there may be a difference in detail method of contact and mode switching, we consider that there is a common mechanism throughout various living creatures in regulating system.
On that basis, we recognized that a question to solve is whether the system to know global properties is necessary for the proportion in response to a whole situation.
Even today, however, the mechanism on proportion regulation from purely local contacts has not been focused and elucidated.

Our view of regulating mechanism in nature is the creature-likeness in behavior, which change its own mode against other behaviors, in addition to an idea of the kinetic theory of molecules in chemical reactions.
This study investigates a simple but plausible mechanism on self-organized proportion regulation between two modes using a simple cellular automata model, as a result from a series of local contacts between individual behaviors.
The most essential question is whether a global information is required for global response, and we found that a study using a continuous model cannot distinguish that. 
A discrete system is crucial in proportion regulation mechanism.
Simulation results with cellular automata model show that the system can appear global responses without any global information, taking the observation data on cellular slime mold as an example.
Finally, we discuss an idea for the mechanism of regulation in multiple task, which can be applied for control of a swarm of self-driven robots. 

\section*{Methods}
\ \ \ \ 
To investigate the mechanism on self-organizing proportion regulation, we confirm the following three presuppositions as clues for modeling: there are no leader in a group; the total number of individuals is conserved; each individual has its own mode, e.g., prestalk or prespore in cellular slime mold, and can switch it into another one.
In addition, given the previous researches~\cite{Brown-Firtel1999,Kay1983,Inoue1989,Gordon2011,Hojo2015,Hayashi2015}, this study consider the two assumptions; each individual can know the mode information of the others through contact; each individual cannot acquire global information, including group size and density.

\subsection*{Concept for regulation mechanism}
\begin{figure}[t]
\centerline{\includegraphics[width=.5\textwidth]{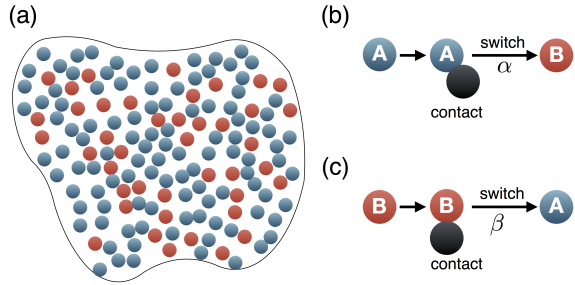}}
\caption{{\bf Conceptual schemes of model.} (a) A swam of individuals with boundary. Mode switching (b) from mode A to mode B and (c) from mode B to mode A, which are performed stochastically after appropriate contact with other individuals. Variables $\alpha$ and $\beta$ are probabilities of mode switchings. \label{fig:concept}}
\end{figure}

\ \ \ \ 
From the presuppositions mentioned above, the system to deal with is a swarm of individuals with boundary, and the total number of individuals is constant, as shown in Fig~\ref{fig:concept}a, where each individual is expressed by a circle.
Each individual has either two kinds of internal modes.
Hence, this study let them mode A and mode B, and assume that each mode can be switched into another one by contact with a few neighborhood others.
In contact between two individuals, three combinations of the modes are recognized, i.e., A-A, A-B (or B-A), and B-B.
After gain the mode information of the others through contact with others, an individual mode is changed with probability- or threshold-based.
In this study, probability-based investigations will be performed, but threshold-based mode switching is possible to discuss similarly.
That case, count the number of the situation which satisfies the required conditions for mode switching.

The required conditions for mode switching can be considered to be dependent on others' mode who physically contacted.
Here, as an example, we consider the following four conditions for mode switching through physical contact and acquisition of the information of others' mode as shown in Figs~\ref{fig:concept}b and c.
\begin{enumerate}
\item Even if an individual with mode A contacts with an individual that has the same mode (mode A), any mode switching will not occur.
\item If an individual with mode A contacts with an individual that has another mode (mode B), will switch its own mode into mode B with probability $\alpha$ (Fig \ref{fig:concept}b).
\item If an individual with mode B contacts with an individual that has the same mode (mode B), will switch its own mode into mode A with probability $\beta$ (Fig \ref{fig:concept}c).
\item Even if an individual with mode B contacts with an individual that has another mode (mode A), any mode switching will not occur.
\end{enumerate}
Probabilities of the mode switching $\alpha$ and $\beta$ are constant, respectively, and would be determined with gene level in actual phenomena.
The important thought behind the concept is creature-likeness in individual behavior.
Each rule seems to describe ``synchrony behavior", ``imitating behavior", ``repulsive behavior", and ``insensitive behavior", respectively. 
These actions are widely observed in living creatures including human beings, therefore, this study shows these are essential for order formation through communication in social animals.
Note that this study would not discuss a detailed method of contact with others and mode switching.

\subsection*{Monte Carlo algorithm for simulation of cellular automata model}
\ \ \ \
In order to realize the concept of contact and mode switching, we demonstrate a two-dimensional ($L\times M$) cellular automata model that adheres to the conservation rule using the Monte Carlo method.
At first, each individual has the mode either A or B, initial positions of the individuals are randomly selected, and the initial numbers of individuals with each mode, $a_0$ and $b_0$, are appropriately established using condition $a_0 + b_0 = N$, where $N$ is the total number of individuals.
In one step, an individual is chosen randomly and moves to a place where no one exists among the four adjacent cells (random walk with the excluded volume effect, EVE).
After the movement, one obtains the mode information of adjacent four cells (the von Neumann Neighborhood).
If the information satisfies the condition of the mode switching, one changes its own mode depending on the probability $\alpha$ or $\beta$.
Note that individuals in adjacent four cells do not change its mode because one who can change its mode is only one individual in one step.
On the boundary, random walks and contact with others for mode switching are limited due to reflecting boundary condition. 

\section*{Results}
\subsection*{Demonstration by Cellular Automata Model}
\begin{figure}[t]
\centerline{\includegraphics[width=.5\textwidth]{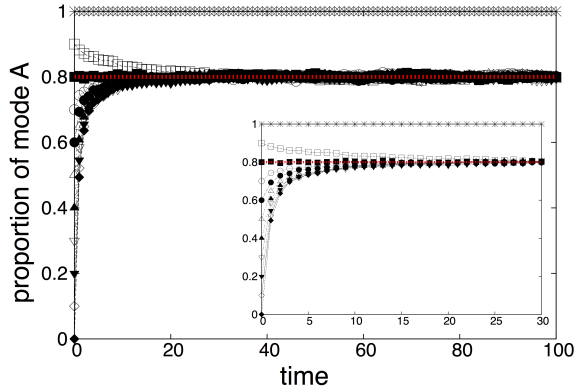}}
\caption{{\bf Calculation results on the proportion of mode A with respect to time.} Horizontal axis is Monte Carlo time step.
Broken line (red in color) is the theoretical value, 0.8.
Parameters are set to $\alpha = 0.2$, $\beta = 0.8$, $ L = M = 100$, and $N=8000$.\label{fig:002}}
\end{figure}

\ \ \ \
Fig~\ref{fig:002} shows the calculation results by cellular automata model in varying initial value $a_0$ in a square domain ($L = M = 100$) and a high density ($N = 8000$).
The proportions of mode A individuals always converge over time to significantly close values, $0.8$, except the case when all individuals initialize with mode A ($a_0 = N$).
The converged value is clearly originated from probabilities of mode switchings (for details, see S1 Appendix).
In this case, $\alpha$ and $\beta$ set to $0.2$ and $0.8$, respectively. 
Here, remarkable results are the time and the frequency of mode switching necessary for convergence of the proportion value. 
Fig~\ref{fig:002} indicates that a few contacts with less than $40$ individuals is enough for each individual to achieve the perfect proportion regulation despite existence of $8000$ individuals in a group.
Note that one time step in Monte Carlo simulation indicates the time step that all individuals averagely are given a chance to contact with others.
In this case, one Monte Carlo time step equals $8000$ simulation steps.

\begin{figure}[t]
\centerline{\includegraphics[width=.5\textwidth]{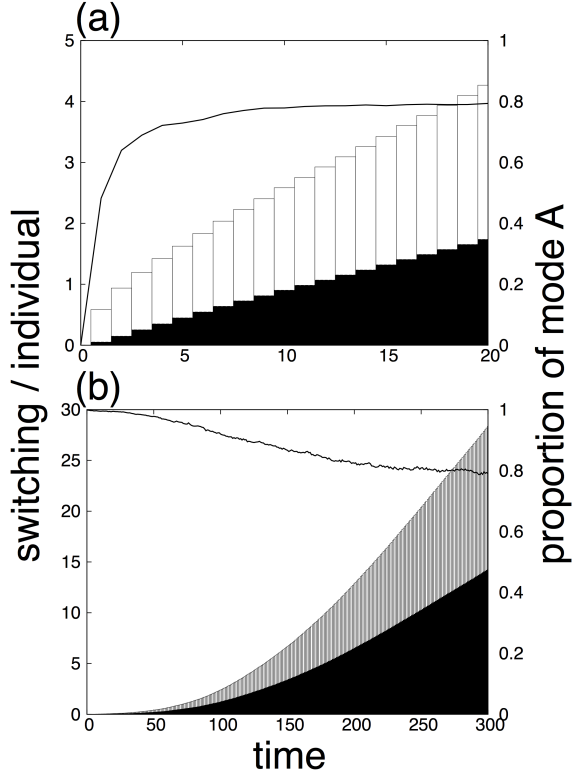}}
\caption{{\bf Calculation results on average frequency of mode switching per each individual.} 
Open and filled boxes are average frequency of mode switching from mode A to B and the opposite direction, respectively.
Solid line is proportion of mode A individuals with respect to Monte Carlo time step.
Initial proportions of mode A individuals are (a) $0$ and (b) $99.875$ percent, respectively.
Parameters are set to $\alpha = 0.2$, $\beta = 0.8$, $ L = M = 100$, and $N=8000$.\label{fig:003}}
\end{figure}

The required contact frequency for convergence depends on the initial proportion.
Fig~\ref{fig:003}a, which is the calculation result in the case of the initial condition $a_0=0$, shows that the proportion get close to convergence value by at most $20$ times.
In this period, actually, mutual mode switchings are performed about four times per an individual.
It has found that this mechanism is a very efficient and reality-based, because the proportion as a whole group can be regulated by contacts with few others nearby ($0.25$ percent of the total) and just four times mode switchings ($20$ percent of the contacts).
In the case that the initial proportion of $a_0$ is extremely high, however, it takes much time for regulation.
For example, Fig~\ref{fig:003}b shows that about $300$ times and $30$ mode switching are required for regulation in the case of the initial condition $a_0=7990$ ($99.875$ percent of the total).
Nevertheless, it is sufficiently-small and efficient compared with the situation that each individual has to interact with all of a group.

\begin{figure}[t]
\centerline{\includegraphics[width=.7\textwidth]{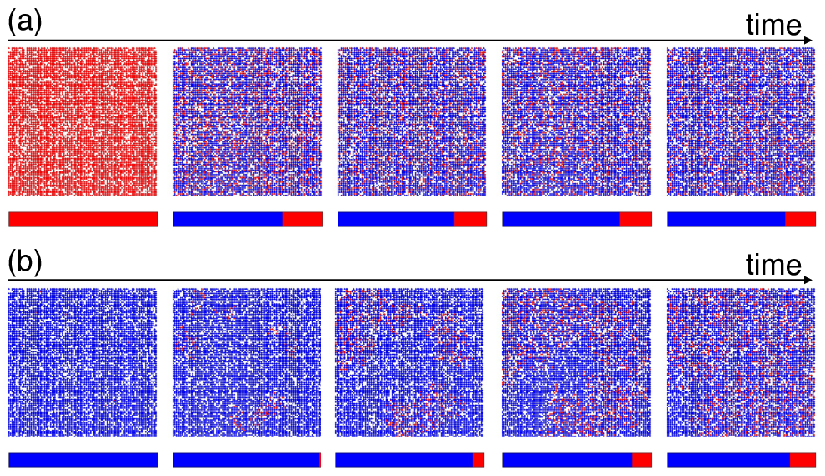}}
\caption{{\bf Simulation results of cellular automata model.}
Blue and red dots are individuals of modes A and B.
Initial proportions of mode A individuals are (a) $0$ and (b) $99.875$ percent, and each panels are every (a) $5$ and (b) $50$ time steps, respectively.
Each bottom gage expresses the percentage of populations of both mode A and mode B individuals.   
Parameters are set to $\alpha = 0.2$, $\beta = 0.8$, $ L = M = 100$, and $N=8000$.\label{fig:004}}
\end{figure}

These differences of taken times can be discussed by the spatial features of mode switching.
The panels of Fig~\ref{fig:004} are the simulation results with time evolution of the case of Fig~\ref{fig:003}, respectively.
Mode switching from mode B to mode A is performed spatial at random (Fig~\ref{fig:004}a), hence the taken time tends to be short.
In contrast, the regions where include mode A individuals expand gradually as shown in Fig~\ref{fig:004}b because the opposite direction of mode switching is required the contact between individuals with two kinds of modes. 

\begin{figure}[t]
\centerline{\includegraphics[width=.4\textwidth]{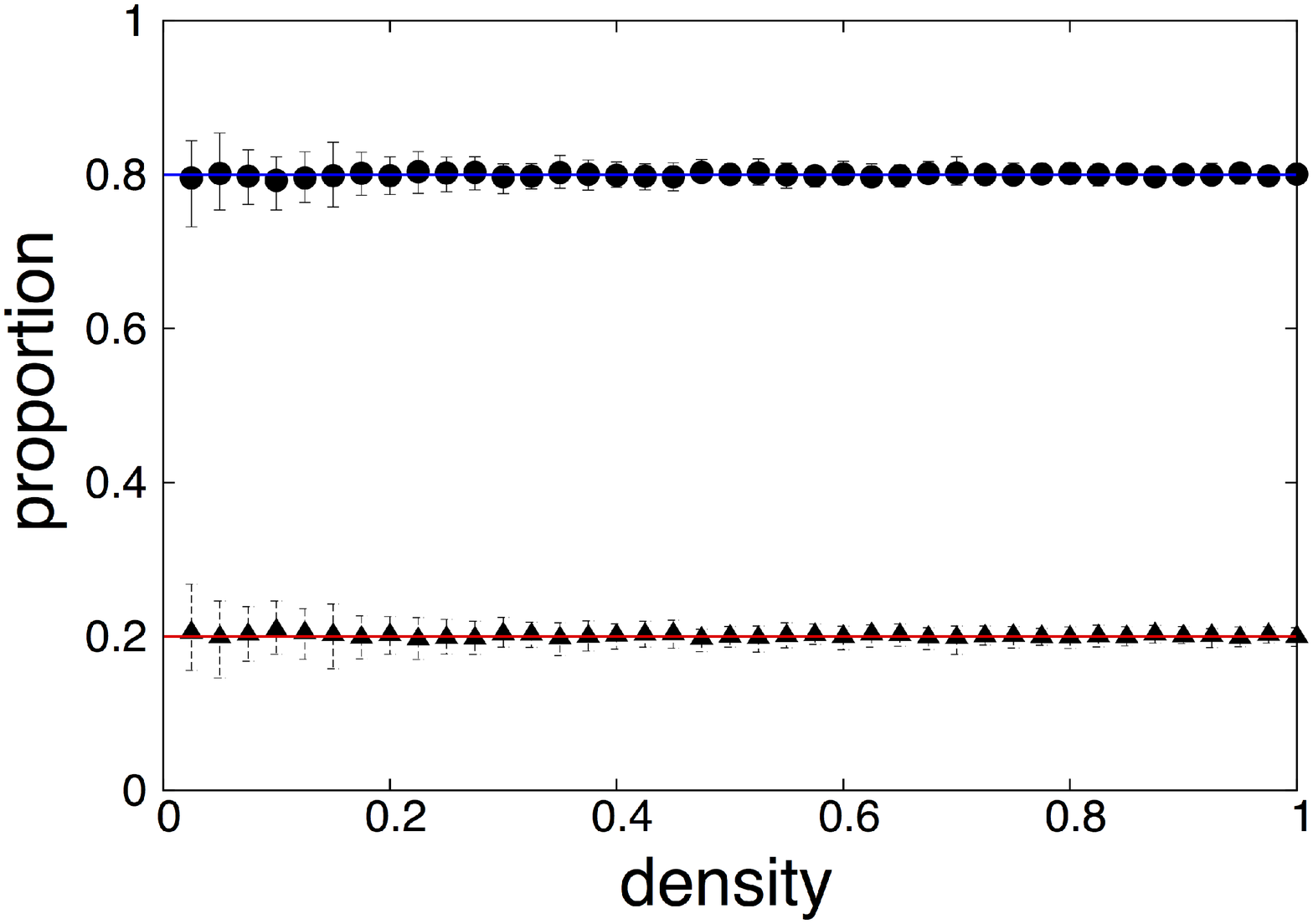}}
\caption{{\bf Simulation results of converged proportions versus density in symmetric case $(m, n) = (1, 1)$}
Circles and triangles are mode A and mode B individuals when $a_0 = 0$, respectively.
Parameters set to $L=M=100$, $\alpha = 0.2$ and $\beta = 0.8$. 
Blue and red lines plot the theoretical solution $0.8$ and $0.2$, respectively. 
\label{fig:005}}
\end{figure}

That exemplifies perfectly a robust mechanism for regulation of desired proportion since it is independent of the global properties, i.e., density, as shown in Fig~\ref{fig:005}.
In addition, we confirmed that random walk is not required in the case of high density in order to regulate the proportion that results from the local contact between two individuals.

\subsection*{Density dependence as a result of local interactions}
\ \ \ \ 
As was clearly observed in the previous experimental researches~\cite{Nanjundiah1995,Oohata1995,Pruitt-Goodnight2014}, proportions may be regulated in response to current overall conditions such as density and group size.
To investigate the origin of these global-dependent features, here consider the more real situations that the contacts between two individuals can be successively performed in nature.
If this chain of contacts is sufficiently fast compared with the mode switching period, it can essentially be viewed as an contact between multiple individuals.
Hence, the second and third rules of mode switching can be generalized as follows. 

 \begin{itemize}
\item[2*.] If an individual with mode A contacts with $m$ ($\geq 1$) individuals who have another mode (mode B), it will reverse its mode with probability $\alpha$, imitating the others' behavior (mode B).
\item[3*.] If an individual with mode B contacts with $n$ ($\geq 0$) individuals who have the same mode (mode B), it will exchange its mode for the opposite one (mode A) with probability $\beta$, repelling the others' behavior.
 \end{itemize}
 
\begin{figure}[h]
\centerline{\includegraphics[width=.5\textwidth]{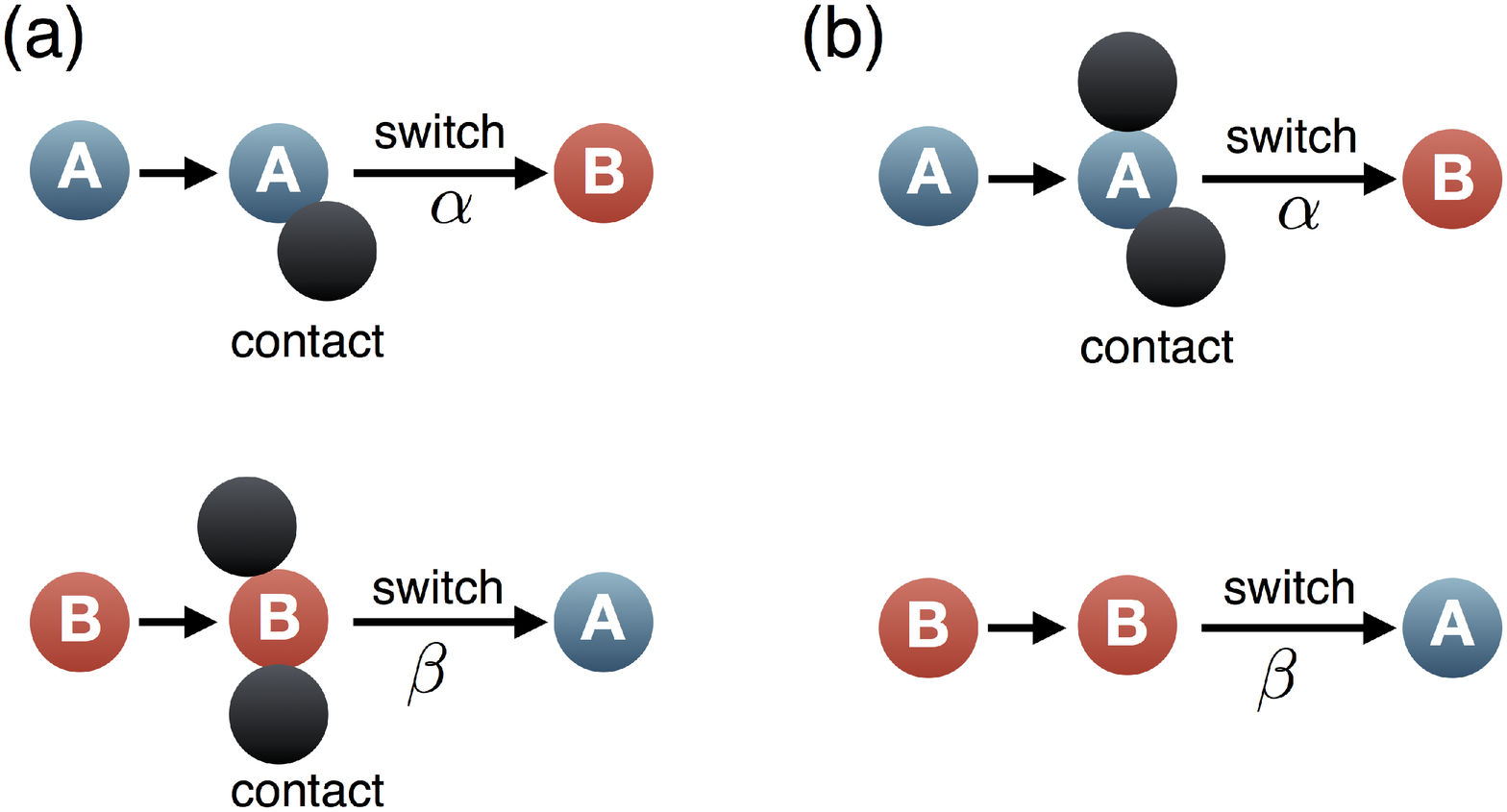}}
\caption{{\bf Conceptual schemes of asymmetric contacts.}
Mode switchings with (a) $(m,n) = (1, 2)$ and (b) $(m,n) = (2, 0)$, where $\alpha$ and $\beta$ are probabilities of mode switchings.\label{fig:006}}
\end{figure}

\begin{figure}[h]
\centerline{\includegraphics[width=.5\textwidth]{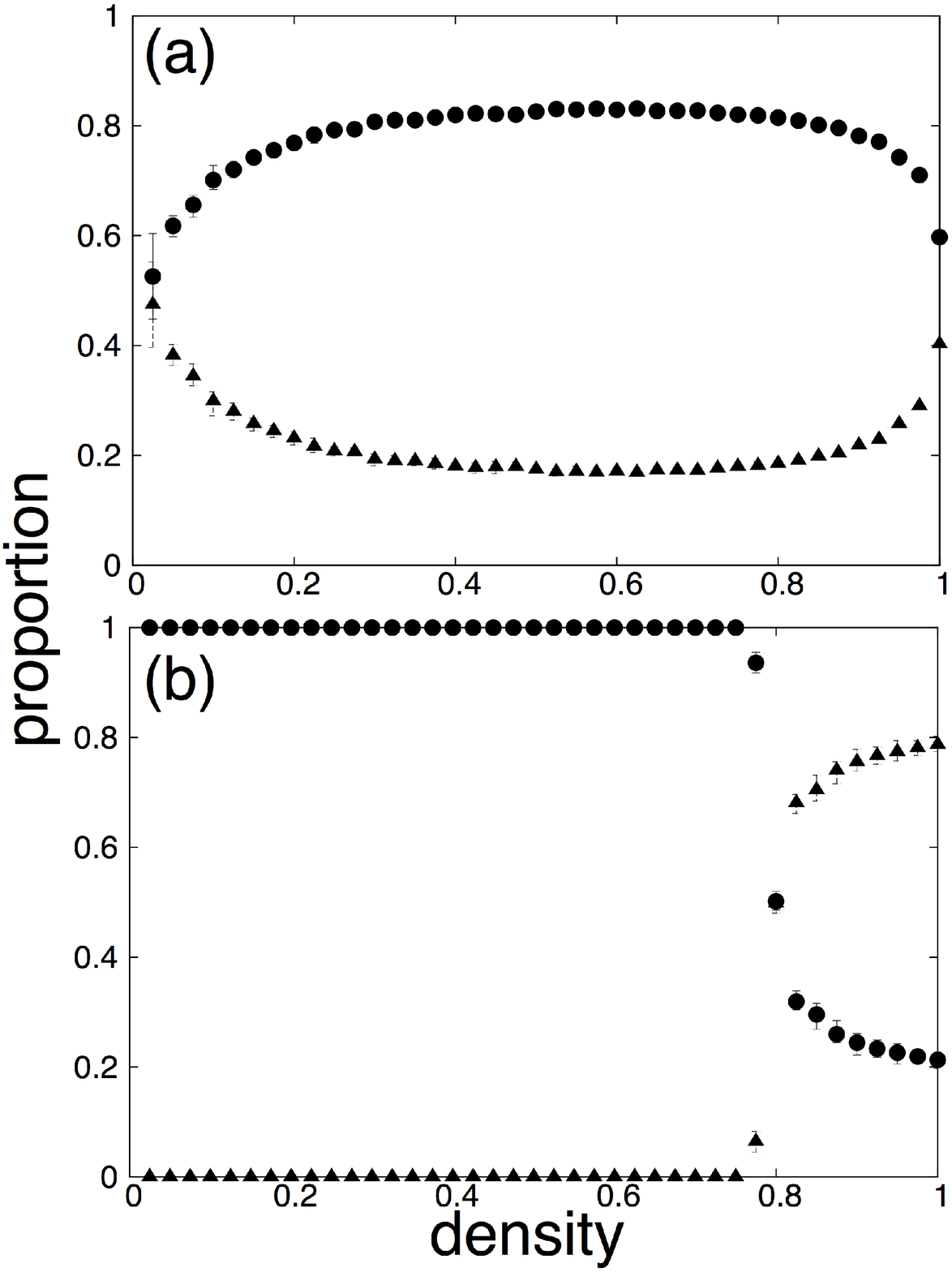}}
\caption{{\bf Simulation results of converged proportions versus density in asymmetric cases.}
Circles and triangles are mode A and mode B individuals when $a_0 = 0$, respectively.
(a) Case $(m,n) = (1, 2)$. Parameters set to $\alpha = 0.01$ and $\beta = 1.0$.
(b) Case $(m,n) = (2, 0)$. Parameters set to $\alpha = 0.8$ and $\beta = 0.2$.
In all simulations, $L=M=100$.\label{fig:007}}
\end{figure}

Both $m$ and $n$ are integer numbers, and either should be greater than or equal to one in accordance with the assumption pertaining to the contacts between individuals.
In demonstrations of cellular automata model, $m$, $n$ are limited to less than or equal to four because of the restrictions of a von Neumann neighborhood.
The above results (Figs~\ref{fig:concept}, \ref{fig:002}, \ref{fig:003} , \ref{fig:004} and~\ref{fig:005}) reflect the case in which $(m, n) = (1, 1)$ and indicates that a group maintains its absolute proportion regardless of the global properties, i.e., density, of a symmetric $(m = n)$ scenario.
Conversely, in the asymmetric cases, i.e., $m \neq n$, proportions are not always unique.
This study demonstrates the two cases, $(m, n) = (1, 2)$ and $(2, 0)$, for instance, as shown in Figs~\ref{fig:006}a and~\ref{fig:006}b, respectively.
We confirmed that the proportion converges to an approximate constant and does not depend on the initial composition of the individuals in the $(1, 2)$ case (see Fig~\ref{fig:s1}(a)).
This result is similar to the symmetric case.
On the other hand, the $(2, 0)$ case has multiple equilibrium proportion values (see Fig~\ref{fig:s1}(b)).
Hence, the system has two stable proportions, depending on the initial composition (for details, see S2 Appendix).
Herein we mention that both cases clearly appear density-dependent.
As shown in Fig~\ref{fig:007}a, the proportion of mode A individuals in the case $(1, 2)$ tends to decrease with a decrease in the density.
In the region of high density, it also slightly decreases with an increase of the density.
Furthermore, in most part of density low side of the case $(2, 0)$, the proportion is regulated into the situation that all individuals are mode A (Fig~\ref{fig:007}b).
The proportion of mode A individuals decreases with a increase of density when density exceed a certain threshold value. 

These results suggest that a reduction in the frequency of contacts, which is implicitly varied by the global properties, affects the proportion.
A decrease in encounters can result from a lower density.
Given a lower density, the unbalanced rates of mode switchings, which result from the difference in values $m$ and $n$, are easily affected because of the lack of information interchange.
That is a result that contacts that require fewer numbers are easily performed.
By contrast, the proportion regulation may be influenced by the excluded volume effect in the high density region.
Detailed analysis on density-dependent are mentioned in the discussion section.

\section*{Discussion}
\ \ \ \
The proposed simplest model that uses local interactions could explain the multiple proportion regulations, which are adapted to current situations.
This strategy's most crucial advantage is its lack of required global information for each individual.
In addition, the model has many variations although it is simple.
Hence, it has the potential to sufficiently model various biological phenomena.
In particular, Fig~\ref{fig:007} indicates that the previous observations, which are recognized as global responses, can be realized only using local information.
As can be seen in chemical equilibrium constant in reversible chemical reaction, the most simple regulating system of proportion requires no contact with others ($m=n=0$).
However, living creatures had acquired two advantages to survive by communication.
The first is an adaptive regulation in response to global property with temporal change.
The second is, which may be often overlooked, that they can sustain the status quo of its internal mode during no contact with anyone.

Before beginning the discussion compared with the previous experimental studies, we mention about the concept of mode switching again.
The simulations, in this study, were performed with stochastic-based mode switching, but it may be difficult in real situations to deal with the required condition of mode switching as a stochastic subject.
We therefore have already confirmed that threshold-based mode switching can realize similar results as the probability-based ones with appropriate values of $\alpha$ and $\beta$ (e.g., $\alpha = 4$ and $\beta =1$ in the case $(1,1)$), which are the number of the situation satisfied the required conditions.

\begin{figure}[h]
\centerline{\includegraphics[width=.5\textwidth]{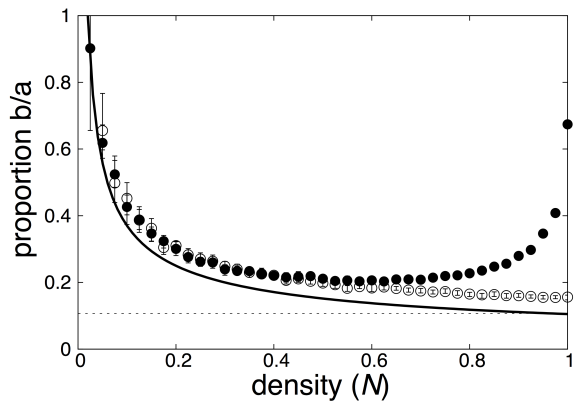}}
\caption{{\bf Simulation results versus density and theoretical solution versus $N$.}
Filled circles are plotted the simulation results of fully-converged proportions $b/a$ with the excluded volume effect (EVE) and open circles are the one without EVE.
Parameters set to $L=M=100$, $\alpha = 0.01$ and $\beta = 1.0$.
Solid line is the theoretical solution $b/a$ versus $N$ in chemical reaction kinetics.
Broken line is the simulation result without spatial constraint.\label{fig:008}}
\end{figure}

Here we discuss examples about the proportion regulation observed in nature.
Simple examples involving two kinds of modes is the spore-stalk proportion in cellular slime mold {\it D. discoideum}, which has reported that the proportions of spores and stalks cells were roughly constant, $80$ to $20$, regardless of the size of fruiting bodies~\cite{Raper1940}. 
The more careful studies have found out that smaller fruiting bodies tend to have higher percentage of stalks (Figs 1 and 2 in~\cite{Nanjundiah1995}).
The puzzling fact for the experimenters was that larger fruiting bodies also have a slightly higher percentage of stalks (Fig 2a in~\cite{Stenhouse-Williams1977}).
Interestingly, the clear tendency as if a cell aggregate knows its size and regulates the proportion according to it can be observed in the simulation result (Fig~\ref{fig:007}a).
Now, we provide a theoretical perspective about the origin of the tendency in the proportion.
We found that the two facts of the stalk proportions, which are higher in both smaller and higher size, result from different origins.
The proposed concept and the rules of mode switchings can be regarded as chemical reversible reaction, and describe with differential equations from chemical reaction kinetics.
From stability analysis of the differential equations, it was clear that the convergence values of the proportion depend on the total amount $N$ in the case of asymmetric contacts (for details, see S2 Appendix). 
In Fig~\ref{fig:008}, which is the case $(1, 2)$, the solid line is a theoretical solution of the differential equations of the ratio $b/a$ with respect to $N$.
On the other hand, filled circles are plotted the numerical result using cellular automata model with EVE with respect to density, and are in qualitatively agreement with the theoretical one in low density region.
In the region of high density, EVE affects on the proportion, as evidenced the comparison with open circles (without EVE).
This is derived just from the creature-likeness that has a volume, it would be impossible to try to understand by the differential equations.
Additionally, the broken line is a case of performing numerical calculations without spatial constraint (in this case, independent of the density), we recognized that a bit of difference between the numerical results and the theoretical value is a discretization error.

Other experiments in cellular slime mold show some facts, which may support our concept. 
The prestalk proportion in the other wild-type strain (Fig 1 in~\cite{Oohata1995}) has a similarity to the simulation result of the case $(2, 0)$ (Fig~\ref{fig:005}b).
In addition, the proportion regulation in smallest fruiting bodies, which were pointed out that cannot understand by differential equations such as reaction-diffusion system~\cite{Bonner2008}, can be realized by our system.
On the other hand, discussing the specifics of the properties of our system, a new mode A individual (switched from a mode B individual) can be generated without other mode A individual, but a new mode B individual cannot be generated without other mode B individual.
That is, for realizing adaptive proportion regulation, the mode B region always has to include the mode A a little bit.
It seems similar that a few anterior-like region remains in posterior region through sorting and migration in the slug of cellular slime mold (Fig 3 in~\cite{Bonner1999}).
It means that the posterior region cannot generate any prestalk cell if the posterior region is consist of prespore type only.
Hence, the remaining anterior-like region is necessary to survive in a dangerous situation (such a cutting experiment~\cite{Raper1940}).
Particularly, cellular slime mold {\it D. discoideum} has been studied as a model organism for cell differentiation and multicellular organisms, this study is expected to be applied to such areas.

This study views the mechanism of proportion regulation in ants colony differently from the reaction threshold model~\cite{Theraulaz1996}, which is generally accepted today.
Although the model can realize a constant proportion with external perturbations, two main questions had been remained.
One is how each ant can gain the information of colony size, because the model including the total number of ants in the colony to realize the proportion in response to the change of colony size. 
In addition, the recent study~\cite{Hayashi2015} that ants keep changing at short intervals its own worker mode, active or inactive, raised a question about the essential factor of the reaction threshold model that each ant has its own characteristic threshold against work.
This study provides a simple answer to these issues.
Our results clearly indicate that we do not need to answer to the question of the method to get the information about the colony size, because the mechanism on local interactions can respond to the change of the colony size without its size information.
Moreover, temporal switching between active and inactive work modes, supports our theory.
The mode switching from inactive to active workers (and vice versa) is caused by the contacts with other ants rather than the reaction against workload.  
If we have a discussion about individual specificity of ants, it would be of the responsiveness to communication, which is described as probability of mode switching $\alpha$ and $\beta$ in our model, not of the reactivity to the amount of work.

Task allocation within the colonies of ants seems to be complex because of the various tasks, but the fact that workers can change a work place with their own age~\cite{Wilson1990,Robinson1992,Gordon2005}  and that an exchange of information is performed between two individuals~\cite{Gordon2011,Hojo2015} supports our concept.
We confirmed that our concept of two mode regulation can apply to multiple mode regulation by considering independently multiple reversible mode switchings (see Fig~\ref{fig:s3}).
In addition, the combination of different densities, which exists within a given colony, i.e., inside (high) or outside (low) the nest, causes various proportions. 
Transitions between tasks occur in concert within the current colony composition~\cite{Gordon1989}.
Thus, transitions can occur in the direction that buffers external perturbation~\cite{Gordon1995} by changing the inside and outside populations~\cite{Gordon2005}.
Hence we expect that our concept will reveal a mechanism on the phenomena in colonies of social insects, which has not been understood. 

Although the mechanism that allows individuals to understand a global situation remained a mystery for many years, it is not mysterious if we consider the fact that individuals do not require any global information in order to have the necessary knowledge.
This study was inspired by chemical reaction kinetics, but chemical reactions are generally described using continuous models, and individual molecules are stochastically averaged.
However, in the study of living organisms, discreteness is crucial.
On the other hand, although self-organization models that apply pattern formation have been proposed~\cite{Turing1952, Keener-Tyson1986}, the phenomena that occur without patterns have not been addressed.
However, the key to understanding living organisms lies within the behaviors that do not exhibit patterns.
We believe that this study brought in renewed questions in the field of proportion regulation in various phenomena in nature since we have recognized that the similar results on proportion with size dependency may be realized by the different causes.
In other words, we does not suggest that all proportion regulations occur as a direct result of local information.
As a primitive research, it requires to confirm which proportion regulation phenomena use global information, given the fact that proportions can be regulated without the use of global information.
In the phenomena without global information, detailed method of local communication would be elucidated in the next step.

Furthermore, this study provides a primary mechanism for self-organized proportion regulation that applies to controls in teeming computers and robotics.
In the modern world of technology and science, we determine a strategy that controls teeming computers and robotics, i.e., task allocation.
Because the existing top-down control is limited given vast amounts of devices and data, a self-organized control method will most likely become the pervasive method of control in the near future.
However, no method exists for the fully self-organization of discretized devices.
Given the advantages of this model, it can directly apply to the control methods used in the field of robotics.
Although each individual requires only a small amount of intelligence, the collective group still acquires the knowledge necessary to cohesively function.
This is called collective intelligence.

\section*{Supporting Information}

\subsection*{S1 Appendix}

\subsubsection*{Reversible chemical reaction}

\ \ A mode switching through a contact with other individuals can be regarded as a reversible chemical reaction between two kinds of substances A and B.
When concentrations of substances are low enough, activity could be considered as its concentration.
In closed system, the total amount of substances $N$ is conserved $a + b = N$, where $a$ ($> 0$) and $b$ ($> 0$) are the concentrations of the substances A and B, respectively.
The change rate of each reaction at a time $t$ is the same, so we can describe these reactions by chemical reaction kinetics as follows: 
\begin{align}
  \left\{
  \begin{array}{lcr}
        a_t &=& -f(a, b),\\
        b_t &=&  f(a, b),
  \end{array}
  \right.
  \label{m0}
\end{align}
where index $t$ is temporal differentiation.
Summation of both equations is $(a + b)_t = 0$ that means the mass conservation.
In this study, the cellular automata model can be perceived as the reversible reactions that either-or substance acts as a catalyst in both reactions.
We demonstrated the case that a catalyst is the substance B as follows:
\begin{align}
 \begin{array}{ccc}
 {\rm A} +  m{\rm B} &\overset{\alpha}\longrightarrow&  {\rm B} +m{\rm B}\\
  {\rm B} + n{\rm B} & \overset{\beta}\longrightarrow&  {\rm A}+ n{\rm B}
  \end{array}
\end{align}
where $\alpha$ and $\beta$ are the reaction rate constant in each reaction, respectively. 

The case of $m = n = 0$ means that exchange of the two modes occurs without a contact between each individual.
It is nothing short of the simplest reversible chemical reaction, A $\rightleftarrows$ B, and its equilibrium constant $K$ is defined as $K =$b/a$= \alpha/\beta$. 

\subsubsection*{Stability Analysis on Symmetric Contacts}
\ \ This study is focused on a contact between individuals, so either $m$ or $n$ must be greater than or equal to $1$.
When $n = m$, these two reactions can be considered to have a symmetry.
Function of time variation of a simplest example $n = m = 1$ is described as follows: 

\begin{equation}
f(a,b)=  \alpha ab - \beta b^2.
\label{m1}
\end{equation}

\begin{figure}[h]
\centerline{\includegraphics[width=.7\textwidth]{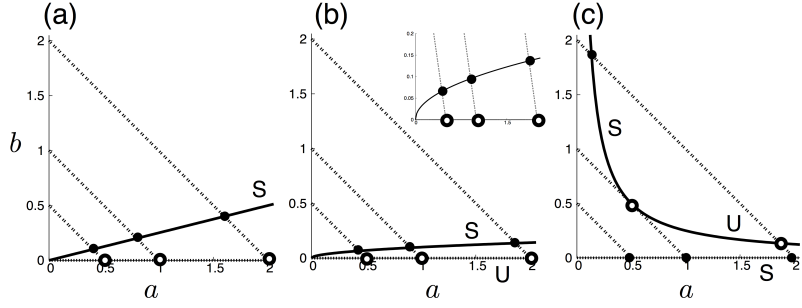}}
\caption{{\bf Phase diagram of differential equations.}  Each amount converges to the intersection between the stable nullcline and the equation of conserved system $a+b=N_0$. (a) Nullcline of the equation (S3) in S1 Appendix with $\alpha = 0.2$ and $\beta = 0.8$, i.e., $b =(\alpha/\beta)a$ (stable) and $b=0$ (unstable).
(b) Nullcline of the equation (S4) in S2 Appendix with $\alpha = 0.01$ and $\beta = 1.0$, i.e., $b =\sqrt{(\alpha/\beta)a}$ (stable) and $b=0$ (unstable).
(c) Nullcline of the equation (S5) in S2 Appendix with $\alpha = 0.8$ and $\beta = 0.2$, i.e., $b =\beta/(\alpha a)$ (stable when $b>a$ and unstable otherwise) and $b=0$ (stable). \label{fig:s2}}
\end{figure}

\begin{figure}[h]
\centerline{\includegraphics[width=.4\textwidth]{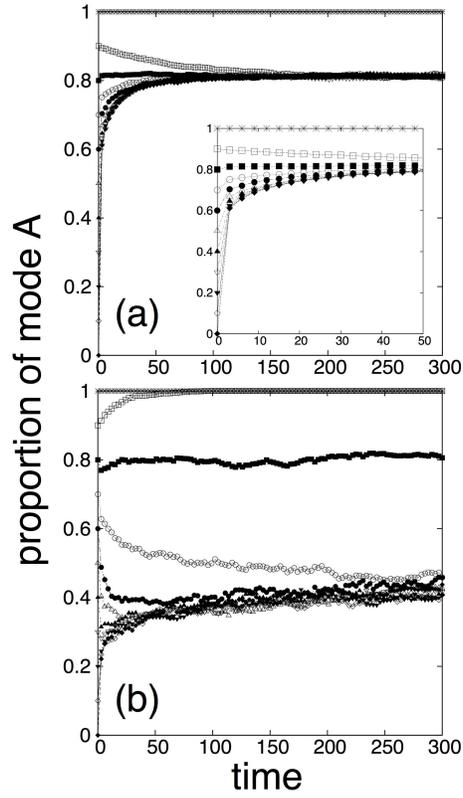}}
\caption{{\bf Simulation results of proportion versus time.} Simulation results of the proportion of mode A individuals with varying the initial proportion, with respect to Monte Carlo time step in (a) the case $(1, 2)$ with $\alpha = 0.01$ and $\beta = 0.8$, and (b) the case $(2, 0)$ with $\alpha = 0.8$ and $\beta = 0.2$. In all simulations,  $L = M = 100$ and $N=8000$. \label{fig:s1}}
\end{figure}

\begin{figure}[h]
\centerline{\includegraphics[width=.5\textwidth]{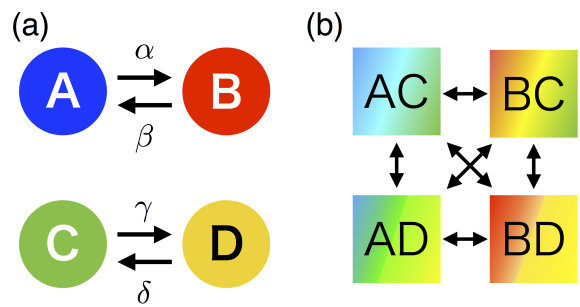}}
\caption{{\bf Conceptual schemes of multiple modes regulation.} (a) Conceptual schemes of independent mode switching between mode A and mode B (top) and between mode C and mode D (bottom), which are performed stochastically after appropriate contact with other individuals. Variables $\alpha$, $\beta$, $\gamma$, and $\delta$ are probabilities of mode switching. 
(b) Conceptual schemes of mode switching between four kinds of internal modes, which are considered as the combinations of two independent modes.
Each individual has internal mode either AC, AD, BC, or BD. 
We confirmed by numerical simulations with cellular automata that the regulation of four modes achieves a constant ratio.\label{fig:s3}}
\end{figure}

Although $(a^*, b^*) = (N, 0)$ and $(\beta N/(\alpha+\beta), \alpha N/(\alpha+\beta))$ are the equilibrium solutions, $(N, 0)$ is unstable (see Fig~\ref{fig:s2}(a)).
Hence, in this system, the proportion converges to
\begin{align}
\frac{b}{a} = \lambda \ \ \text{(constant).} \nonumber
\end{align}
where $\lambda = \alpha/\beta$, regardless of initial concentration of mode A $a_0$ except for $a_0 = N$.
In the simulation in the main text, the proportion of model A converges to $0.8$ since $\alpha=0.2$ and $\beta=0.8$.

\subsection*{S2 Appendix}
\ \ Here gives a mathematical explanation for global dependency in our simulation results, which is the asymmetric cases ($n \neq m$).
In this study, we demonstrated two simple examples, $(n, m) = (1, 2)$ and $(2, 0)$.

\subsubsection*{Flexible Proportion : Monostable System}
\ \ In the case of $(n, m) = (1, 2)$, the simulation indicates that the proportion converges uniquely independent of the initial composition as shown in Fig~\ref{fig:s1}(a), but depends on the whole density $N$ as shown in Fig 7a of the main text. 
The density-dependency is explained by the following function for the equation (S1) in the text of S1 Appendix: 

\begin{equation}
   f(a,b) = \alpha ab - \beta b^3.
  \label{m2}
\end{equation}
as shown in S2 Fig (b), $(a^*, b^*)=(N,0)$ is unstable and 
\begin{align}
\left( \frac{\alpha+2\beta N -\sqrt{\alpha^2+4\alpha\beta N}}{2\beta}, \frac{-\alpha +\sqrt{\alpha^2+4\alpha\beta N}}{2\beta} \right)
\end{align}
is stable.
Hence, the proportion converges to
\begin{align}
\frac{b}{a} = \frac{\lambda + \sqrt{\lambda^2+4\lambda N}}{2N}.\nonumber
\end{align}
where $\lambda = \alpha/\beta$, regardless of initial concentration of mode A $a_0$ except for $a_0 = N$.
It found clearly that the proportion depends on total amount $N$.

\subsubsection*{Flexible Proportion : Bistable System}
\ \ In this section, we focused on the system $(n, m) = (2, 0)$.
The function for the equation (S1) in the text of S1 Appendix are as follows: 

\begin{equation}
     f(a,b) =   \alpha ab^2 - \beta b,
  \label{m3}
\end{equation}
as shown in Fig~\ref{fig:s2}(c), $(a^*, b^*) = (N, 0)$ would be only one stable point when $N$ is smaller than a threshold $2\sqrt{\beta/\alpha}$.
When $N$ is greater than the threshold, the system is bistable.
If the initial proportion of mode B individuals is more than the unstable proportion, the proportion converges to 
\begin{align}
\frac{b}{a} = \frac{\lambda+\sqrt{\lambda^2-4\lambda N}}{2}-1.\nonumber
\end{align}
So, the converged proportions depend on the initial proportion and total amount $N$.

\subsubsection*{Origin of global-dependency}

\ \ As indicated above, the converged proportions depend on total amount $N$ in the asymmetric case of contacts ($m \neq n$).
We have to note that two different interpretations are possible for realization of  this size dependency in this differential equation.
The one is by local interaction in a discrete system as shown in Figs 7 and 8 in the main text.
In this case, global dependency appears without a global information for each individual. 

Another is by acquisition of global information for each individual.
Using the conservation property $u+v=1$, the non-dimensional differential equations of the case $(1,2)$ as follows,
\begin{equation}
\left\{
 \begin{array}{clc}
u_{\tau} &=& -\lambda uv + Nv^3,\\
v_{\tau} &=& \lambda uv - Nv^3,
  \end{array}
  \right.
  \label{nd2}
\end{equation}
and of the case $(m,n)=(2,0)$ as follows, 
\begin{equation}
\left\{
 \begin{array}{clc}
 u_{\tau} &=& -\lambda N uv^2 + \frac{v}{N},\\
v_{\tau} &=& \lambda N uv^2 - \frac{v}{N}.
  \end{array}
  \right.
  \label{nd3}
\end{equation}
From analysis of non-dimensional continuous equations, it means that each coefficient has a feature depending on $N$, and it seems that each individual knows a global feature.
Exactly, in the previous studies, the mistake of interpretation of the origin of global-dependency had been induced by such an analysis of continuous models.
Namely, the global-dependence of proportion seems to be realized by only acquisition of global information.

\end{document}